\documentclass[twocolumn,showpacs,preprintnumbers,amsmath,amssymb,amsfonts,pra]{revtex4-1}
\usepackage{graphicx,epsfig}
\usepackage{dcolumn}
\usepackage{bm}

\begin{document}

\title{Spectral transformations in the regime of pulse self-trapping in a nonlinear photonic crystal}

\author{Denis V. Novitsky}
\email{dvnovitsky@tut.by} \affiliation{B.I. Stepanov Institute of
Physics, National Academy of Sciences of Belarus,
Nezavisimosti~Avenue~68, 220072 Minsk, Belarus}

\date{\today}

\begin{abstract}
We consider interaction of a femtosecond light pulse with a
one-dimensional photonic crystal with relaxing cubic nonlinearity in
the regime of self-trapping. By use of numerical simulations, it is
shown that, under certain conditions, the spectra of reflected and
transmitted light possess the properties of narrow-band
(quasi-monochromatic) or wide-band (continuum-like) radiation. It is
remarkable that these spectral features appear due to a significant
frequency shift and occur inside a photonic band gap of the
structure under investigation.
\end{abstract}


\maketitle

\section{Introduction}

The need for taking into account the noninstantaneousness of
nonlinear response of a medium was realized soon after the rise of
nonlinear optics. From the end of the 1960s specialists studied the
influence of nonlinearity relaxation in the framework of the Debye
model on such effects as laser beam self-focusing \cite{Fleck,
Hanson} and parametric amplification \cite{Trillo}. Among recent
studies, attention has been attracted to modulational instability
effects in media with noninstantaneous nonlinearity \cite{Shih,
Velchev, Zhang}, resulting in generation of pulse trains
\cite{SotoCrespo} and solitons \cite{Cambournac}, instability of
speckle patterns \cite{Skipetrov}, and reshaping of solitary pulses
\cite{Liu}.

However, the theoretical nonlinear optics of photonic band gap
materials usually deals with instantaneous processes of
nonlinearity. Many results may be found in reviews and monographs
(see, for example, Refs. \cite{Sakoda, Gaponenko, Bert, Notomi}). We
should also note some effects connected with ultrashort pulse
interaction with nonlinear photonic crystals, such as pulse
compression and temporal soliton formation \cite{Eggl, Zhelt},
subdiffractive propagation \cite{Stal, Loiko}, and pulse
localization on a defect \cite{Good, Mak1, Mak2}.

In this paper we consider spectral transformations of femtosecond
pulses interacting with a one-dimensional photonic crystal with
relaxing cubic nonlinearity. As was shown in our previous
publication \cite{Novit}, light self-trapping occurs in such a
nonlinear structure due to formation of a nonlinear dynamical cavity
(or trap) inside it. The present paper is a logical continuation of
that paper. The importance of spectral investigation is connected
with the possibility of spectral broadening, which in some extremal
cases can result in supercontinuum generation. This phenomenon can
be observed, for example, in photonic crystal fibers \cite{Dudley}
or in filamentation processes in bulk materials \cite{Couairon}.
Spectral broadening is one of the main points of our research.

The paper is logically divided into several sections. In Sec.
\ref{ste} we give the problem formulation and consider some
additional details of the self-trapping effect important for the
present paper. Section \ref{spec} is devoted to the spectral
features connected with the nonlinear interaction of a pulse with a
photonic crystal in the regime of self-trapping. Finally, Sec.
\ref{concl} contains a short conclusion.

\section{\label{ste}Self-trapping effect}

Propagation of an ultrashort pulse in a one-dimensional nonlinear
photonic crystal [a structure of $(AB)^N$ type] is described by the
Maxwell wave equation
\begin{eqnarray}
\frac{\partial^2 E}{\partial z^2}&-&\frac{1}{c^2} \frac{\partial^2
(n^2 E)}{\partial t^2} = 0, \label{Max}
\end{eqnarray}
with the dependence of refractive index on light intensity $I=|E|^2$
as follows,
\begin{eqnarray}
n(z,t)=n_0(z)+\delta n (I, t). \label{refr}
\end{eqnarray}
Here $E$ is the electric field strength, $n_0(z)$ is the linear part
of the refractive index varying along the $z$ axis, and $\delta n$
is the nonlinear part of the refractive index, which is governed by
the Debye model of relaxing nonlinearity \cite{Akhm}
\begin{eqnarray}
t_{nl} \frac{d \delta n}{d t}+ \delta n=n_2 I, \label{relax}
\end{eqnarray}
where $n_2$ is the Kerr nonlinear coefficient and $t_{nl}$ is the
relaxation time, which is assumed to be of the order of several
femtoseconds (fast electronic cubic nonlinearity). Further we
consider femtosecond light pulses with the amplitude of Gaussian
shape $A=A_m \exp(-t^2/2t_p^2)$, where $t_p$ is the pulse duration.
To analyze the interaction of such a pulse with a nonlinear photonic
crystal, we use the finite-difference time-domain method of
numerical simulations which was described in detail in Ref.
\cite{Novit}. The spectra of pulses (incident, reflected,
transmitted) in this paper are calculated as the absolute values of
the Fourier transform of the corresponding field profiles. The
spectra are normalized to the peak value of the incident pulse
spectrum which is recognized as unity.

The parameters used in our calculations are as follows: the linear
parts of the refractive indices of the layers $A$ and $B$ of the
photonic crystal $n_a=2$ and $n_b=1.5$, respectively; their
thicknesses $a=0.4$ and $b=0.24$ $\mu$m; the number of layers
$N=200$; the pulse duration $t_p=30$ fs; the central wavelength of
the initial pulse spectrum is $\lambda_c=1.064$ $\mu$m if not stated
otherwise. The nonlinear coefficient of the material is defined
through the nonlinear term of the refractive index, so that $n_2
I_0=0.005$; this means that the pulse amplitude is normalized by the
value $A_0=\sqrt{I_0}$. The relaxation time of the nonlinearity of
both layers is $t_{nl}=10$ fs.

\begin{figure}[t!]
\includegraphics[scale=0.95, clip=]{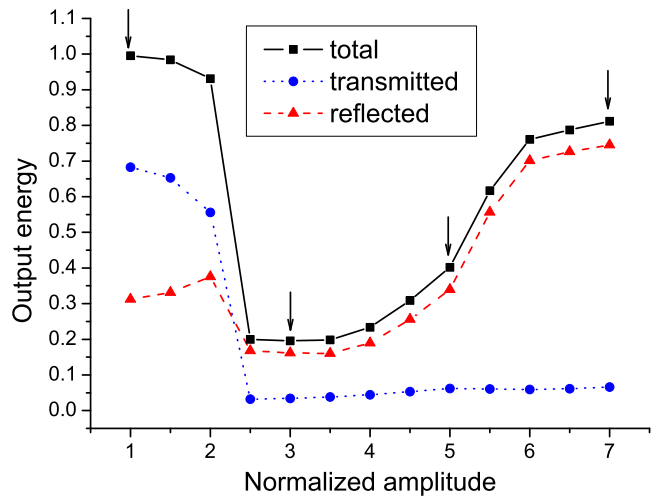}
\caption{\label{fig1} (Color online) Dependence of the output light
energy (normalized to the input energy) on the peak amplitude of the
incident pulse. Energy is integrated over the time $200 t_p$ (about
six times larger than the pulse transmittance time in the linear
regime).}
\end{figure}

As it was predicted in our previous work \cite{Novit}, the
interaction of the pulse (whose duration is comparable to the
relaxation time) with a nonlinear photonic crystal results in the
effect of pulse self-trapping. This situation when the energy of
radiation leaving the structure is only a small fraction of the
incident pulse energy is shown in Fig. \ref{fig1}. It is seen that,
for large enough intensity of the pulse, the output energy
demonstrates a profound decrease corresponding to the self-trapping
of the pulse inside the photonic crystal. The output energy is
calculated by intensity integration over time at the input and
output points of the structure; thus we obtain the energies of the
reflected and transmitted light, and the total output energy is
their sum. As the amplitude of the incident pulse increases further,
light is trapped closer and closer to the input face of the crystal
so that the reflected radiation energy gets larger and larger.
Finally, the trapping occurs near the very input, so that most of
the light is immediately reflected. The range of pulse durations and
relaxation times for the self-trapping effect to be observed in our
structure was studied in Ref. \cite{Novit} as well: $t_{nl}$ varies
from a fraction of a femtosecond to about $150$ fs, and $t_p$ from
about $10$ fs to about $200$ fs.

\begin{figure}[t!]
\includegraphics[scale=0.85, clip=]{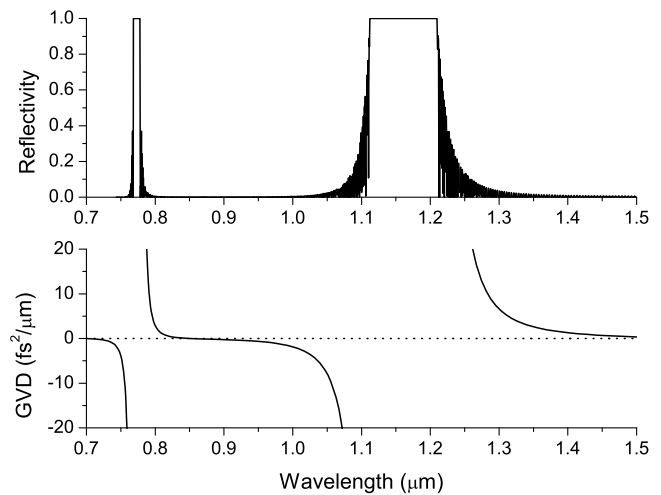}
\caption{\label{fig2} Spectral curves for reflectivity (upper panel)
and group velocity dispersion (lower panel) of the photonic crystal
under consideration. The parameters of the structure are given in
the text.}
\end{figure}

\begin{figure}[t!]
\includegraphics[scale=0.9, clip=]{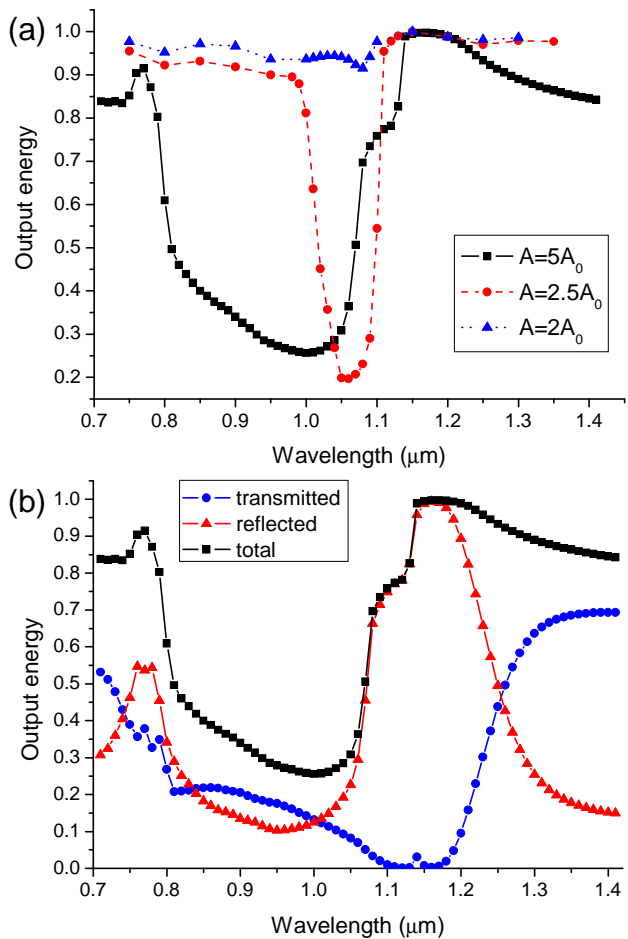}
\caption{\label{fig3} (Color online) Dependence of the output light
energy (normalized to the input energy) on the central wavelength of
the incident pulse. (a) The results for different values of the
amplitude. (b) Comparison of transmitted and reflected energies for
the pulse with $A_m=5A_0$. Energy is integrated over the time $200
t_p$.}
\end{figure}

For better understanding of this effect, let us consider its
frequency dependence. In Fig. \ref{fig2} we see the reflectivity and
group velocity dispersion (GVD) of the photonic crystal considered
as functions of the light wavelength. It is well known that the GVD
parameter $k_2=d^2 k / d \omega^2$ is decisive in observation of
pulse compression \cite{Akhm, Zhelt}. Indeed, if the nonlinearity
coefficient $n_2$ is positive (this is the case in our
consideration), then one needs to have a medium with negative GVD.
Since the self-trapping effect is characteristic for the regime of
pulse compression \cite{Novit}, we can expect that the pulse will be
trapped inside the photonic crystal if the pulse spectrum lies in
the negative dispersion domain. This expectation is justified in
Fig. \ref{fig3} where the output energy dependence on the central
wavelength of the pulse spectrum is represented. The dip in this
dependence is unambiguously correlated with the negative-GVD region
in Fig. \ref{fig2}(b). Moreover, the appearance of a minimum in the
output energy implies that there are some competitive processes
which come in contact and determine the result of the pulse-crystal
interaction. It seems natural to suggest that these processes are
dispersion spreading and nonlinear light-matter interaction as in
the case of usual pulse compression.

Figure \ref{fig3}(a) also shows that, for larger values of the pulse
amplitude, the optimal value of the central wavelength is situated
further from the band gap (compare the curves at $A_m=2.5 A_0$ and
$5A_0$). At the same time, the dip for $A_m=2.5 A_0$ is deeper,
which is in accordance with the assumption about optimal (or close
to optimal) trapping in this case (see Fig. \ref{fig1}). Note that
for $A_m=2 A_0$ there is no self-trapping behavior but,
nevertheless, there is a very shallow dip even closer to the
forbidden gap than at $A_m=2.5 A_0$.

\section{\label{spec}Spectral features of light in the self-trapping regime}

\begin{figure}[t!]
\includegraphics[scale=0.8, clip=]{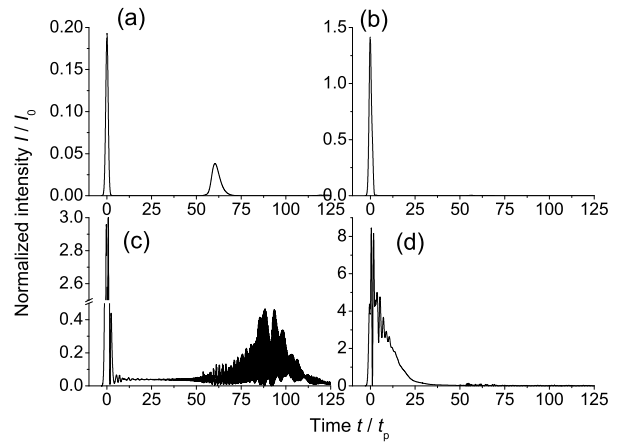}
\caption{\label{fig4} Shape of the reflected intensity at different
peak amplitudes of the incident pulse: (a) $A_m=A_0$, (b)
$A_m=3A_0$, (c) $A_m=5A_0$, and (d) $A_m=7A_0$. The central
wavelength of the pulse spectrum is $\lambda_c=1.064$ $\mu$m.}
\end{figure}

\begin{figure}[t!]
\includegraphics[scale=0.85, clip=]{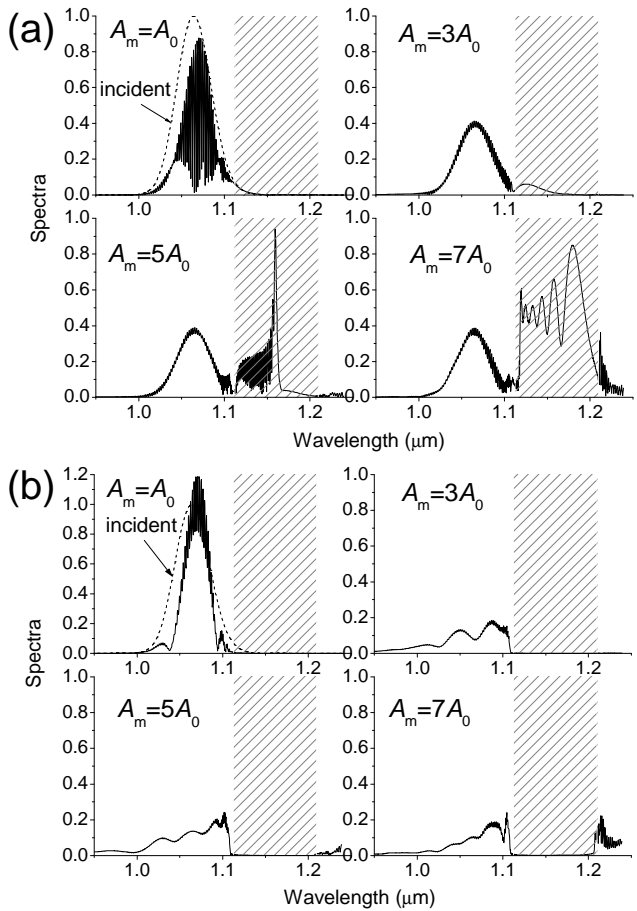}
\caption{\label{fig5} Spectra of (a) reflected and (b) transmitted
radiation at different peak amplitudes $A_m$ of the incident pulse
corresponding to those used in Fig. \ref{fig4}. The dashed curve
depicts the spectrum of the incident Gaussian pulse. The band gap of
the linear photonic crystal is shaded.}
\end{figure}

\begin{figure}[t!]
\includegraphics[scale=0.85, clip=]{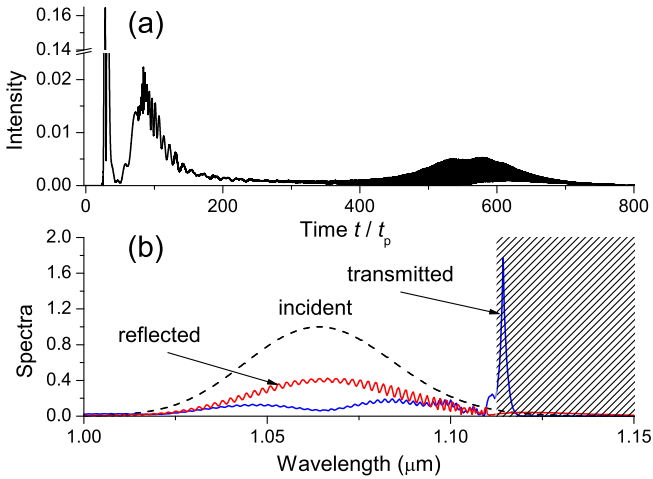}
\caption{\label{fig6} (Color online) (a) Shape of the transmitted
intensity. (b) Spectra of reflected and transmitted radiation. The
peak amplitude of the incident pulse is $A_m=3A_0$. Calculations
were performed for the photonic crystal with linear $A$-layers
(first layers of the period). The band gap of the linear photonic
crystal is shaded.}
\end{figure}

\begin{figure}[t!]
\includegraphics[scale=0.85, clip=]{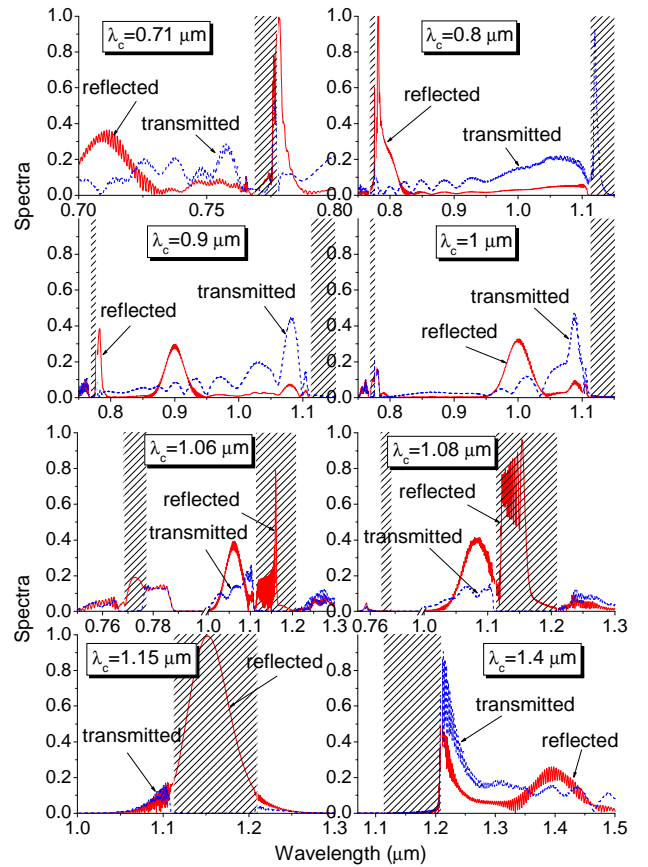}
\caption{\label{fig7} (Color online) Spectra of reflected and
transmitted radiation at different central wavelengths $\lambda_c$.
The peak amplitude of the incident pulse is $A_m=5A_0$. The band gap
of the linear photonic crystal is shaded. The position of the
spectrum of the incident pulse is characterized by the bell-shaped
curve in the spectrum of the reflected light.}
\end{figure}

\begin{figure*}[t!]
\includegraphics[scale=0.85, clip=]{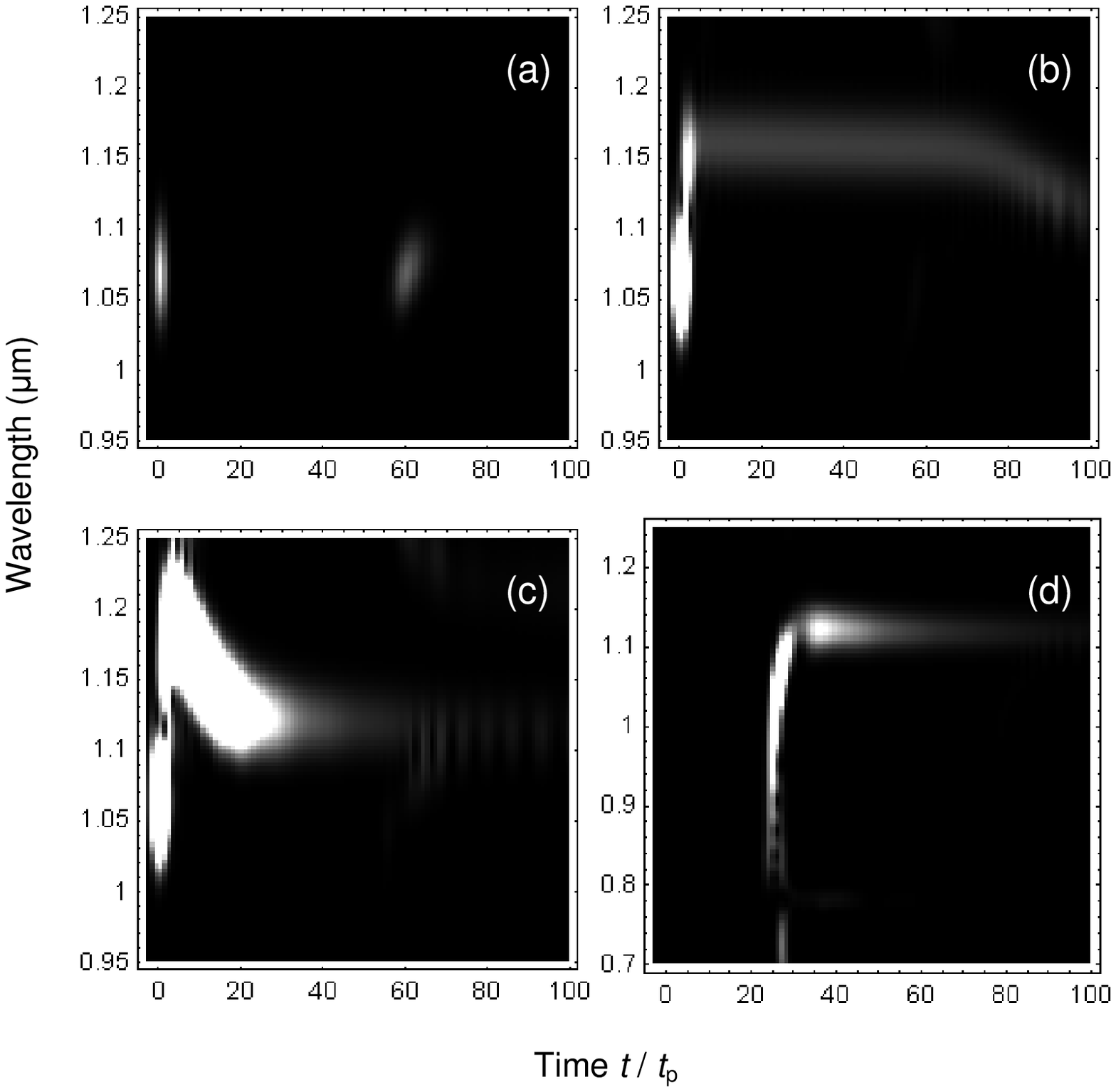}
\caption{\label{fig8} (a)-(c) Spectrograms of reflected radiation at
peak amplitudes of the incident pulse $A_m=A_0$, $5A_0$, and $7A_0$,
respectively. The central wavelength is $\lambda_c=1.064$ $\mu$m.
(d) Spectrogram of transmitted radiation at $A_m=5A_0$ and
$\lambda_c=0.8$ $\mu$m.}
\end{figure*}

The change in shape of the reflected intensity connected with the
dynamical trap formation can be traced in Fig. \ref{fig4}, where we
plotted the results for different values of incident pulse amplitude
corresponding to the cases marked by the arrows in Fig. \ref{fig1}.
Most of all, we are interested in Figs. \ref{fig4}(c) and
\ref{fig4}(d), which show some remarkable features to be discussed
in detail. In contrast to the usual peaks of reflected light in
Figs. \ref{fig4}(a) and \ref{fig4}(b), corresponding to the cases of
zero trapping and maximal trapping, respectively, Fig. \ref{fig4}(c)
was calculated for a point on the upward slope of the reflected
energy curve (the amplitude is $A_m=5A_0$). This means that
self-trapping still exists but some part of the radiation leaves the
nonlinear trap inside the photonic crystal. As one can see in Fig.
\ref{fig4}(c), this leaving radiation represents almost stationary
radiation for quite a long time. As a result, we can expect that the
spectrum of reflected light has to possess a pronounced narrow peak
corresponding to this quasi-monochromatic radiation. This
expectation is entirely justified, as spectral plots demonstrate in
Fig. \ref{fig5}(a). It is seen that, for $A_m=5A_0$, the spectrum of
reflected light really has a sharp peak, while in the cases
$A_m=A_0$ and $A_m=3A_0$ the spectra approximately correspond to the
spectrum of the incident Gaussian pulse. Moreover, the position of
the spectral peak also provokes our interest, since it is situated
deep inside the photonic band gap of the structure.

If we take a pulse with greater amplitude ($A_m=7A_0$), then, as
mentioned above, the reflected light appears immediately with a wide
and seemingly unstructured envelope [Fig. \ref{fig4}(d)]. It turns
out that its spectrum in this case completely covers the band gap in
continuum-like fashion [see Fig. \ref{fig5}(a)]. Note that the
spectra of transmitted radiation do not intrude into the forbidden
gap, as is witnessed by Fig. \ref{fig5}(b). In general, spectral
broadening can be linked with self-phase-modulation resulting in the
generation of new frequencies in the pulse spectrum due to the
temporal variation of the refractive index \cite{Couairon}. However,
the fact that the sharp edges of the spectrum include just the
entire band gap seems to be unexpected. In fact, we have a situation
when light converts under nonlinear interaction in such a way that
the spectrum is more and more pulled into the forbidden gap. On the
other hand, we should keep in mind that self-trapping is connected
with a local change of reflective properties of the photonic crystal
\cite{Novit}.

Thus, if we are on the upward slope of Fig. \ref{fig1}, we can
obtain narrow-band radiation in reflection. In other words, this
corresponds to stronger coupling between the pulse and the nonlinear
structure than in the case of the optimal self-trapping effect due
to the greater value of the incident intensity. One might suggest
that something similar should be observed in the opposite situation
when the light-medium interaction gets weaker, i.e. in the region of
the very abrupt downward slope in Fig. \ref{fig1}. Obviously, the
narrow-band spectrum is expected to be obtained in transmitted (not
reflected) radiation in this case. In order to prove this statement,
we use another method to make the coupling between the pulse and the
nonlinear photonic crystal weaker. We take materials with smaller
nonlinearity, rather than decreasing the intensity of the pulse. In
Fig. \ref{fig6} the results are shown for the $(AB)^N$-structure
with linear $A$ layers, while the parameters of the $B$ layers
remain unchanged. It is seen that the structure of the transmitted
radiation is similar to that of Fig. \ref{fig4}(c). As a result, in
the spectrum of transmitted light a pronounced quasi-monochromatic
peak occurs [Fig. \ref{fig6}(b)]. This peak, however, is situated
near the very edge of the band gap of the linear photonic crystal.
Obviously, low-intensity quasi-monochromatic radiation seen in Fig.
\ref{fig6}(a) cannot significantly change the refractive properties
of the structure through which it is to be transmitted. Therefore,
there is only a slight shift of the forbidden gap, which can be
referred to as a self-induced transparency effect in the nonlinear
photonic crystal.

Let us return to Fig. \ref{fig3} and consider spectral
transformations for incident pulses with different central
wavelengths $\lambda_c$. The dip corresponding to the self-trapping
phenomenon is situated between two band gaps plotted in Fig.
\ref{fig2} (we call it the inter-gap region). Changing $\lambda_c$
in this region, one can obtain all the variants of spectral
peculiarities discussed above and even more as can be seen in Fig.
\ref{fig7}. Further, we list the main features seen in this figure:

(i) When the spectrum of the incident pulse is out of the inter-gap
region ($\lambda_c=0.71$ $\mu$m, which is the region of negative
GVD; see Fig. \ref{fig2}), the spectrum of reflected radiation has a
sharp peak near the very low-frequency (on the dip side) edge of a
narrow band gap.

(ii) If the spectrum is in the positive-GVD domain ($\lambda_c=0.8$
$\mu$m), the peak for reflected radiation still occurs, but there
also appears a peak in the transmitted light spectrum near the edge
of the wider (long-wave) band gap. Self-induced transparency is also
observed due to the shift of the forbidden gap. Obviously, this
situation corresponds to the weak light-matter coupling regime
discussed previously in connection with Fig. \ref{fig6}.

(iii) As we move further inside the negative-GVD region (the region
of the self-trapping dip, $\lambda_c=0.9$ and $1$ $\mu$m), the peaks
near both the wide and narrow gaps diminish and become less and less
pronounced (the peak for transmitted light moves away from the edge
of the gap).

(iv) At $\lambda_c=1.06$ $\mu$m we see a narrow quasi-monochromatic
peak actually inside the forbidden gap (compare with Fig.
\ref{fig5}, $A_m=5A_0$). Note that the reflected radiation also
appears inside the narrow (short-wave) band gap.

(v) At $\lambda_c=1.08$ $\mu$m the reflected light spectrum in the
band gap widens and takes a continuum-like shape with characteristic
oscillatory fine structure (compare with Fig. \ref{fig5} at
$A_m=7A_0$).

(vi) If the initial spectrum is almost entirely inside the band gap
($\lambda_c=1.15$ $\mu$m), then we have the usual reflection without
any evidence of nonlinear interaction.

(vii) At $\lambda_c=1.4$ $\mu$m we are also outside the dip and
inside the positive-GVD region. The spectra of both reflected and
transmitted light demonstrate a sharp break right on the edge (the
side opposite to the dip) of the band gap. There is no any sign of
self-induced transparency.

Finally, to make clear the connection between temporal curves and
spectra, we turn to the spectrogram technique widely used in
supercontinua investigation \cite{Dudley}. The spectrogram is
calculated as \cite{Dudley}
\begin{eqnarray}
S(\omega, \tau) = \left|\int_{-\infty}^{+\infty} E(t) g(t-\tau)
\textrm{e}^{-i \omega t} dt \right|^2, \label{spectrogram}
\end{eqnarray}
where $E(t)$ is the field under investigation (in our case, the
reflected or transmitted field), $g(t)=\exp(-t^2/2t_p^2)$ is the
gate function which is chosen to be a replica of the input pulse.
The spectrogram $S(\omega, \tau)$ allows an intuitive understanding
of the correlation between temporal and spectral features of a given
signal. In Fig. \ref{fig8} such spectrograms are shown; Figs.
\ref{fig8}(a)-\ref{fig8}(c) corresponding to temporal and spectral
curves of the reflected radiation depicted in Figs. \ref{fig4} and
\ref{fig5}(a), respectively. Figure \ref{fig8}(d) represents the
spectrogram of transmitted radiation of the spectrum demonstrated in
the upper right panel of Fig. \ref{fig7} (the case of
$\lambda_c=0.8$ $\mu$m).

The spectrogram of reflected light at the incident pulse amplitude
$A_m=A_0$ [Fig. \ref{fig8}(a)] shows two intensity peaks seen in
Fig. \ref{fig4}(a). These peaks are concentrated near the central
wavelength $\lambda_c=1.064$ $\mu$m without any significant
frequency shift. Such a shift is easily seen in Fig. \ref{fig8}(b)
at $A_m=5A_0$, so that low-intensity quasi-monochromatic radiation
occurs exactly inside the photonic band gap. It is also worth noting
that at $t \geq 75 t_p$ a frequency shift in the reverse direction
(toward $\lambda_c$) exists. This fact can be associated with the
chaotic ending of quasi-monochromatic radiation seen in Fig.
\ref{fig4}(c). For larger input intensity ($A_m=7A_0$), this reverse
shift occurs earlier in time, but the spectrum covers a wider
frequency range with approximately uniform intensity giving rise to
continuum-like radiation inside the band gap. The last spectrogram
[Fig. \ref{fig8}(d)] shows the transmitted quasi-monochromatic
radiation with a large frequency shift from $\lambda_c=0.8$ $\mu$m
to about $1.12$ $\mu$m. It is easily seen that this shift happens
very fast in time.

\section{\label{concl}Conclusion}

In conclusion, in this paper we have studied the spectral
transformations of ultrashort (femtosecond) light pulses resulting
from their interaction with a nonlinear photonic crystal in the
regime of self-trapping. In our analysis we used only the processes
of light self-action, so that the effect of generation of wide and
narrow spectra cannot be connected by high harmonics and
sum-frequency appearance. However, as our results demonstrate, these
self-interaction processes are sufficient for impressive spectral
transformations in nonlinear photonic crystals. These
transformations concern both reflected and transmitted light spectra
and depend on the regime of light-material interaction. In
particular, if this interaction is strong (the pulse is trapped near
the entrance of the photonic crystal), a narrow peak and
continuum-like spectral features occur in reflected light. On the
other hand, if the light-structure interaction is weak (the pulse is
trapped near the exit of the photonic crystal), a narrow peak near
the edge of the band gap appears in the transmitted light spectrum.
Obviously, relaxing (noninstantaneous) behavior of the nonlinearity
and periodic change of the linear refractive index (a photonic
crystal \emph{per se}) are the key conditions due to the necessity
of self-trapping.

We should also say a few words about perspectives of this research.
First, some improvements are possible in the realization and control
of the self-trapping effect by adjustment of the photonic structure.
In particular, a chirped photonic crystal with varying period can be
employed to shift the trapping position inside the structure and,
perhaps, to relax the requirements on the materials. However, this
modification is still to be studied in detail. The second question
is connected with the possibility of experimental realization of
self-trapping and the corresponding spectral effects. Although the
parameters used do not belong to some specific nonlinear medium,
they seem to be quite realistic. The relaxation times (a few
femtoseconds) are characteristic of media with a fast electronic
mechanism of Kerr nonlinearity. However, such media possess
relatively low nonlinear coefficients; therefore one has to use
high-intensity pulses ($\sim 100$ GW/cm$^2$) and take into account
the problem of the damage threshold, which is high enough in the
case of femtosecond pulses. We believe that one can find some
materials (for example, doped glasses \cite{Sutherland}) which
satisfy all these conditions.

\begin{acknowledgements}
The author thanks Dr. Andrey Novitsky for productive discussions and
keen interest in this investigation. The work was supported by the
Belarusian Foundation for Fundamental Research (Grant No. F11M-008).
\end{acknowledgements}


\begin{thebibliography}{26}
\bibitem{Fleck} J.A. Fleck and P.L. Kelley, \apl {\bf15}, 313 (1969).
\bibitem{Hanson} E.G. Hanson, Y.R. Shen, and G.K.L. Wong, \ap {\bf14}, 65 (1977).
\bibitem{Trillo} S. Trillo, S. Wabnitz, G.I. Stegeman, and E.M. Wright, \josab {\bf6}, 889 (1989).
\bibitem{Shih} M.-F. Shih, C.-C. Jeng, F.-W. Sheu, and C.-Y. Lin, \prl {\bf88}, 133902  (2002).
\bibitem{Velchev} I. Velchev, R. Pattnaik, and J. Toulouse, \prl {\bf91}, 093905 (2003).
\bibitem{Zhang} L. Zhang et al., \oc {\bf283}, 2251 (2010).
\bibitem{SotoCrespo} J.M. Soto-Crespo and E.W. Wright, \apl {\bf59}, 2489 (1991).
\bibitem{Cambournac} C. Cambournac et al., \josab {\bf19}, 574 (2002).
\bibitem{Skipetrov} S.E. Skipetrov, \ol {\bf28}, 646 (2003).
\bibitem{Liu} X. Liu, J.W. Haus, and S.M. Shahriar, \oc {\bf281}, 2907 (2008).
\bibitem{Sakoda} K. Sakoda, \textit{Optical Properties of Photonic Crystals}, 2nd ed. (Springer, Berlin, 2005).
\bibitem{Gaponenko} S.V. Gaponenko, \textit{Introduction to Nanophotonics} (Cambridge University Press, New York, 2010).
\bibitem{Bert} M. Bertolotti, J. Opt. A {\bf8}, S9 (2006).
\bibitem{Notomi} M. Notomi, Rep. Prog. Phys. {\bf73}, 096501 (2010).
\bibitem{Eggl} B.J. Eggleton, R.E. Slusher, C.M. de Sterke, P.A. Krug, and J.E. Sipe, \prl {\bf76}, 1627 (1996).
\bibitem{Zhelt} A.M. Zheltikov, N.I. Koroteev, S.A. Magnitskiy, and A.V. Tarasishin, Quantum Electron. {\bf 28}, 861 (1998).
\bibitem{Stal} K. Staliunas, C. Serrat, R. Herrero, C. Cojocaru, and J. Trull, \pre {\bf74}, 016605 (2006).
\bibitem{Loiko} Yu. Loiko, R. Herrero, and K. Staliunas, \josab {\bf24}, 1639 (2007).
\bibitem{Good} R.H. Goodman, R.E. Slusher, and M.I. Weinstein, \josab {\bf19}, 1635 (2002).
\bibitem{Mak1} W.C.K. Mak, B.A. Malomed, and P.L.Chu, \pre {\bf67}, 026608 (2003).
\bibitem{Mak2} W.C.K. Mak, B.A. Malomed, and P.L.Chu, \josab {\bf20}, 725 (2003).
\bibitem{Novit} D.V. Novitsky, \pra {\bf81}, 053814 (2010).
\bibitem{Dudley} J.M. Dudley, G. Genty, and S. Coen, \rmp {\bf78}, 1135 (2006).
\bibitem{Couairon} A. Couairon and A. Mysyrowicz, Phys. Rep. {\bf441}, 47 (2007).
\bibitem{Akhm} S.A. Akhmanov, V.A. Vysloukh, and A.S. Chirkin, \textit{Optics of Femtosecond Laser Pulses} (AIP Press, New York, 1992).
\bibitem{Sutherland} R.L. Sutherland, \textit{Handbook of Nonlinear Optics}, 2nd ed. (Marcel Dekker, New York, 2003).
\end{thebibliography}
\end{document}